\documentclass[journal]{IEEEtran}
\usepackage{amsmath,graphicx}
\usepackage{epstopdf}
\usepackage{color}
\usepackage{wrapfig}
\usepackage{lipsum}
\usepackage{float}
\usepackage{colortbl,booktabs}
\usepackage{mathtools}
\usepackage[compress]{cite}
\definecolor{myRed}{rgb}{0.8, 0.2, 0.2}
\definecolor{myYellow}{rgb}{0.2,0.2,0.8}

\newcommand{\figref}[1]{Fig.~\ref{#1}}

\newcommand{\eqnref}[1]{(\ref{#1})}

\usepackage{amssymb}

\usepackage{multirow}
\usepackage{float}
 \usepackage{bm}
\usepackage{subfigure}
\usepackage{multicol}
\usepackage{algorithm}
 \usepackage{mathrsfs}
\usepackage{bm}
\usepackage{amsthm}
\theoremstyle{definition}

\theoremstyle{Proposition}
\theoremstyle{remark}

\usepackage{cite}
\usepackage{stfloats}%
\usepackage{lettrine}
\usepackage{algpseudocode}
\usepackage{indentfirst,multicol}
\usepackage{amsmath}
\usepackage{cases}
\usepackage{lettrine}
\usepackage[margin=0.48in]{geometry}

\ifCLASSINFOpdf
\else
\fi
\begin{document}

\title{Low-Complexity Signal Detection for the Splitting Receiver Scheme}
\author{Yanyan Wang, \emph{Member, IEEE}, Qidi Li, Xiaohu Tang, \emph{Senior Member, IEEE}}

\maketitle

\IEEEpeerreviewmaketitle
\begin{abstract}
\let\thefootnote\relax\footnote{
The work of Y. Wang was supported by the National Natural Science Foundation of China under Grant 62201476, and in part by the  Natural Science Foundation of Sichuan Province under Grant 2021YJ0549.(\emph{Corresponding author: Yanyan Wang}.)

	
Y. Wang, Q. Li, and X. Tang are with School of Information Science and Technology, Southwest Jiaotong University, Chengdu, China  (email: yanyanwang@swjtu.edu.cn; lqd2020200504@my.swjtu.edu.cn; xhutang@swjtu.edu.cn).

}
This letter proposes a low-complexity signal detection method for the splitting receiver scheme, which achieves an excellent symbol error rate (SER) performance.
Based on the three-dimensional (3D) received signal of the splitting receiver, we derive an equivalent two-dimensional (2D) signal model and  develop a low-complexity signal detection method for the practical modulation scheme.  The computational complexity of the proposed signal detector is reduced by at least 11.1 times as compared to the 3D signal detection scheme shown in~\cite{wang2020on}. Simulation results demonstrate that the approximation SER achieved by the low-complexity signal detection is very close to that of the accurate SER at a certain power splitting ratio.
\end{abstract}

\section{Introduction}
Receiver design has received a amount of attention in wireless communication systems. The existed receiver schemes are generally categorized into two classes: coherent receiver~\cite{proakis2007digital} and non-coherent receiver~\cite{Chowdhury2016Scaling}.
The coherent receiver is based on the coherent detection (CD), which has been extensively utilized in cellular communications and local area networks because of the better performance gain than those of the non-coherent receivers.
Although the power detection (PD) based non-coherent receiver suffers from performance loss compared to the coherent receivers, it can provide the advantages of low power consumption and low cost, which makes it attractive for the internet of things (IoT), backscatter communications\cite{gao2019energy, liu2021en}, etc.



Recently, a new receiver architecture was proposed in~\cite{Liu2017A, wang2020on, wang2022sp, WangGCJ21}. The motivations of the works are to develop a practical radio-frequency (RF) splitting receiver for providing better performance than the traditional non-splitting receiver. For the splitting receiver architecture, the received signal is split into two streams. Then, the two streams separately go through the CD and non-coherent receivers  and thus  obtain the three-dimensional (3D) signals. Ignoring the antenna noise, the work in~\cite{Liu2017A} proposed the splitting receiver architecture. Based on the received signal model, the logarithmic likelihood ratio detection algorithm \cite{kim2012so} was utilized to detect the transmitted signals.  The follow-up work in~\cite{wang2020on} formulated a more practical system model, where both the antenna noise and post-processing noises were considered. The maximum likelihood (ML) detection method was used in~\cite{wang2020on} for recovering the transmitted signals based on the 3D received signal. Since the likelihood  function has no closed-form expression for the splitting receiver scheme, the computational complexity  of the optimal ML detector is very high. To reduce the computational complexity, a relatively low-complexity signal detection method was proposed in~\cite{wang2020on} through deriving the closed-form likelihood function expression. However, the likelihood function expression is still very complicated. Particularly, when the modulation order is large, the signal detection complexity will further increase.

In this letter, we propose a low-complexity signal detection scheme for the splitting receiver architecture and  analyze the symbol error rate (SER) performance.
Specifically,
based on the 3D splitting received signal, we formulate an equivalent  two-dimensional (2D) signal model. As a result, the signal detection is transformed into the minimum distance detection by changing the coordinate system. The complexity of the proposed signal detection method is greatly reduced compared with the  ML detection that based on the 3D received signal.  Precisely, the computational complexity of the proposed signal detection method is reduced by about $11.1$ times  compared to that of the method in~\cite{wang2020on}.
Notably, simulation results show that the SER performance of the low-complexity detector is mostly indistinguishable from that of the optimal detector  at a certain power splitting ratio.


The remainder of the paper is organized as follows. In Section II, the splitting receiver with considering practical noise model is illustrated.  In Section III, the low-complexity detector of the splitting receiver is developed.  Numerical results are provided in Section IV. Section V analyzes the computational complexity of the proposed signal detection scheme. Finally, Section VI concludes the letter.
\begin{figure}[t]
\centering
\includegraphics[width=1.0 \linewidth]{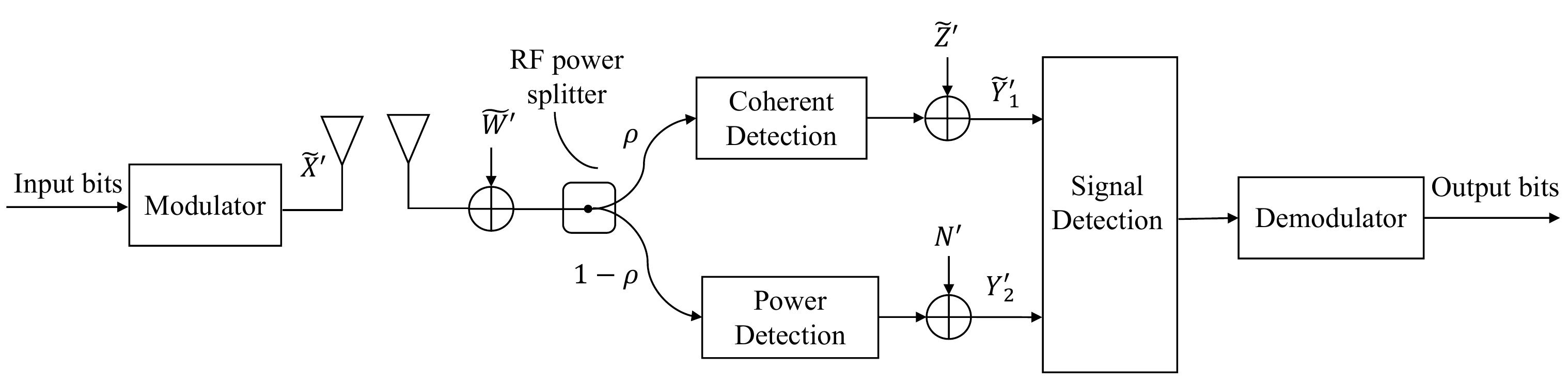}
\caption{The communication system model with splitting receiver architecture.}
\label{fig:Fig1sysmod}
\end{figure}

\section{System Model}
\figref{fig:Fig1sysmod} shows a single-antenna communication system with consideration of the antenna noise $\tilde{W}^{'}$, where a splitting receiver architecture is utilized at the receiver side. The information bits are modulated to generate complex transmitted signals ${{\tilde{X}}}$ with average transmit power $P$ and then propagated through the wireless fading channel $\tilde{h}$.
Assuming that the channel gain is $|\tilde{h}|$ and phase shift of the channel is $\phi$, the equivalent complex channel is written as $\tilde{h}\triangleq|\tilde{h}|e^{j\phi}$.
At the receiver side, the received signal is split into two streams by a power splitter $\rho$. Then, the two streams respectively go through the CD  and PD-based non-coherent receivers.
Finally, the signals from the CD  and PD receivers are jointly used for information detection.

Based on the splitting receiver architecture, the received signals are expressed as
\begin{equation}
\label{equ:CDreceiverc}
{{{\tilde{Y}}}^{'}_1} = \sqrt{\rho}(\sqrt{P}\tilde{h}{{\tilde{X}}}+{{\tilde{W}^{'}}})+{{\tilde{Z}^{'}}},
\end{equation}
\begin{equation}
\label{equ:PDreceiverc}
{{{Y}}^{'}_2} = \eta (1-\rho)|\sqrt{P}\tilde{h}{{\tilde{X}}}+{{\tilde{W}^{'}}}|^2+{{N}}^{'},
\end{equation}
where $\rho \in [0,1]$ and $\eta\in (0,1]$ respectively denote the power splitting ratio and the conversion efficiency, which have been analyzed in~\cite{wang2020on}. Specifically, when $\rho=0$ and $\rho=1$, the splitting receiver is degraded to the PD receiver and the CD receiver, respectively.

For the ease of analysis, let ${{{\tilde{Y}}}_1} = e^{-j\phi}{{{\tilde{Y}}}^{'}_1}$ and ${{{Y}}_2}={{{Y}}^{'}_2}/\eta$. Then, we can obtain $e^{-j\phi}{{{\tilde{Y}}}^{'}_1}=e^{-j\phi}\sqrt{\rho}(\sqrt{P}|\tilde{h}|e^{j\phi}{{\tilde{X}}}+{{\tilde{W}}^{'}})+e^{-j\phi}{{\tilde{Z}}^{'}}$ and ${{{Y}}^{'}_2}/\eta=|e^{-j\phi}|^2(1-\rho)\big|\sqrt{P}{|\tilde{h}|e^{j\phi}}{{\tilde{X}}}+{{\tilde{W}}^{'}}\big|^2+{{N}}^{'}/ \eta$.  Hence, the received signals are rewritten as
\begin{equation}
\label{equ:CDreceiver}
{{{\tilde{Y}}}_1} = \sqrt{\rho}(\sqrt{P}|\tilde{h}|{{\tilde{X}}}+{{\tilde{W}}})+{{\tilde{Z}}},
\end{equation}
\begin{equation}
\label{equ:PDreceiverch}
{{{Y}}_2} = (1-\rho)\big|\sqrt{P}{|\tilde{h}|}{{\tilde{X}}}+{{\tilde{W}}}\big|^2+{{N}},
\end{equation}
where ${\tilde{W}}\triangleq e^{-j\phi}{{\tilde{W}}^{'}}$, $\tilde{Z}\triangleq e^{-j\phi}{{\tilde{Z}^{'}}}$ and ${{N}} \triangleq {{N}}^{'}/ \eta$. Notably,  ${\tilde{W}}$, $\tilde{Z}$ and ${{N}}$ are called as the antenna noise, conversion noise, and rectifier noise, respectively. These noises still meet zero-mean Gaussian distribution, which follow
$\tilde{W}\sim \mathcal{CN}(0,\sigma_{\textrm{A}}^2)$, $\tilde{Z}\sim \mathcal{CN}(0,\sigma_{\textrm{cov}}^2)$, and $N\sim \mathcal{N}(0,\sigma_{\textrm{rec}}^2)$, respectively.

\begin{figure}[t]
\centering
\includegraphics[width=0.9 \linewidth]{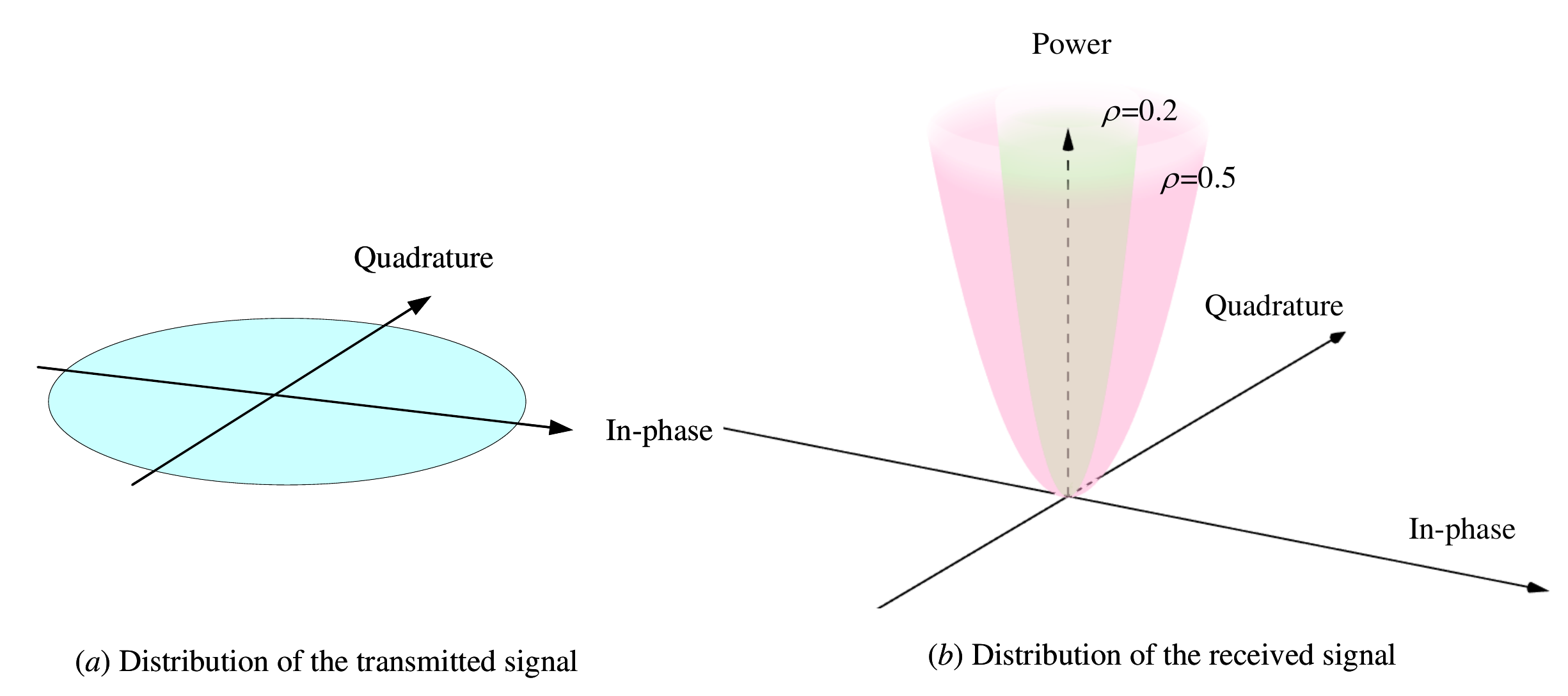}
\caption{Distributions of the transmitted and received signals.}
\label{fig:Fig2sysdis}
\vspace{-0.5cm}
\end{figure}

Considering the $M$-quadrature amplitude modulation (QAM) and $M$-phase shift keying (PSK), the distributions of the transmitted and received signals from \eqnref{equ:CDreceiver} and \eqnref{equ:PDreceiverch} are illustrated in \figref{fig:Fig2sysdis}. It is observed that the constellation points of the transmitted signals $\tilde{X}$ are distributed on the 2D I-Q plane. At the receiver side, the received signals $({{{\tilde{Y}}}_1},{{{Y}}_2})$ are distributed on the 3D  paraboloid in the I-Q-P space when the power splitting ratio $\rho\in(0,1)$. Specifically,  we observe that the received signal space of the traditional coherent ($\rho=1$) and PD-based non-coherent  ($\rho=0$)  receivers are respectively 2D and 1D. Therefore, compared with the traditional receivers, the splitting receiver architecture spreads the received signal space.

\section{ Proposed Low-complexity Signal Detection}
In this section, we project the 3D received signal with high-dimensional space into the 2D signal space and derive an equivalent 2D received signal.
Based on the simplified 2D signal model, a low-complexity signal detection is developed for the practical modulation scheme.
Note that when simplifying the signal model, the $M$-QAM and $M$-PSK schemes are exploited in this letter,  which include the magnitude and phase information for the transmitted signals.

\subsection{Simplified 2D Signal Model}
Without loss of generality, scaling  \eqnref{equ:CDreceiver} by $\sqrt{\rho}\sqrt{P}|\tilde{h}|$  and \eqnref{equ:PDreceiverch} by $(1-\rho)P|\tilde{h}|^2$, the output signal of the splitting receiver is expressed as
\begin{equation}
\label{equ:simCDreceiver}
{{{\tilde{Y}}}_1} = {{\tilde{X}}}+\frac{{{\tilde{W}}}}{\sqrt{P}|\tilde{h}|}+\frac{{{\tilde{Z}}}}{\sqrt{\rho}\sqrt{P}|\tilde{h}|},
\end{equation}
\begin{equation}
\label{equ:simPDreceiverch}
{{{Y}}_2} = \bigg|{{\tilde{X}}}+\frac{{{\tilde{W}}}}{\sqrt{P}|\tilde{h}|}\bigg|^2+\frac{N}{(1-\rho)P|\tilde{h}|^2}.
\end{equation}

From \eqnref{equ:simCDreceiver} and \eqnref{equ:simPDreceiverch}, as the signal power $P$ increases, the 3D received signal $({{{\tilde{Y}}}_1}, {{{Y}}_2})$  can be expressed as the  equivalent 2D signal ${{\tilde{Y}}}$, where received signal
${{{Y}}_2}$  provides the amplitude information, and the received signal ${{{\tilde{Y}}}_1}$ offers the phase information. Defining $\tilde{S}={{\tilde{X}}}+\frac{{{\tilde{W}}}}{\sqrt{P}|\tilde{h}|}$, $\tilde{Z}_s=\frac{{{\tilde{Z}}}}{\sqrt{\rho}\sqrt{P}|\tilde{h}|}$ and $N_s=\frac{N}{(1-\rho)P|\tilde{h}|^2}$, the simplified 2D signal is rewritten as
\begin{align}
\label{equ:sim2D}
{\tilde{Y}}&\triangleq |\tilde{Y}|\textrm{e}^{j\angle{{{\tilde{Y}}}}} \triangleq \sqrt{Y_2}\textrm{e}^{j\angle{{{\tilde{Y}}}_1}}\notag\\
&=\sqrt{|\tilde{S}|^2+{{N}}_s}\textrm{e}^{j\angle( \tilde{S}+{{\tilde{Z}}}_s)},
\end{align}
where $|\tilde{Y}|\triangleq\sqrt{Y_2}$  and $\textrm{e}^{j\angle{{{\tilde{Y}}}}}\triangleq\textrm{e}^{j\angle{{{\tilde{Y}}}_1}}$.
In the practical wireless communication system,  the conversion noise $\tilde{Z}$ is generally much stronger than the rectifier noise $N$~\cite{wang2022sp}.
Therefore, ignoring the noise term ${{N}}_s$ in the high signal-to-noise ratio (SNR) regime, \eqnref{equ:sim2D} is approximated by
\begin{align}
\label{equ:sim2D1}
{\tilde{Y}} &\approx |{{\tilde{S}}}|\textrm{e}^{j\frac{{{\tilde{S}}}+\tilde{Z}_s
}{|{{\tilde{S}}}+\tilde{Z}_s|}} \approx |{{\tilde{S}}}|\textrm{e}^{j\angle {{\tilde{S}}}+j\frac{{
\tilde{Z}}_s}{|{{\tilde{S}}}|}} \notag \\
 &\approx |{{\tilde{S}}}|\textrm{e}^{j\angle {{\tilde{S}}}}\textrm{e}^{j\frac{{{{Z}}}_p}{|{{\tilde{S}}}|}} \notag \\
&={{\tilde{S}}}\left(\cos\left(\frac{{{{Z}}}_p}{|{{\tilde{S}}}|}\right)+j\sin\left(\frac{{{{Z}}}_p}{|{{\tilde{S}}}|}\right)\right),
\end{align}
where ${{\tilde{S}}}=|{{\tilde{S}}}|\textrm{e}^{j\angle {{\tilde{S}}}}$ and ${{{Z}}}_p$ is the projection of ${\tilde{Z}_s}$ to ${{\tilde{S}}}$, which follows ${{{Z}}}_p\sim \mathcal{N}(0,\frac{\sigma_{\textrm{cov}}^2}{2\rho P|\tilde{h}|^2})$.  By adopting  the first-order Taylor-series approximation, \eqnref{equ:sim2D1} is simplified as
\begin{align}
\label{equ:sim2D2}
{\tilde{Y}}
&\approx {{\tilde{S}}}\left(1+j\frac{{{{Z}}}_p}{|{{\tilde{S}}}|}\right)= {{\tilde{S}}}+j{{{Z}}}_p\textrm{e}^{j\angle {{\tilde{S}}}}.
\end{align}
As $P$ increases, it is observed that the argument of the  complex number ${\tilde{S}}$ is almost the same as the ${\tilde{X}}$, i.e., $\textrm{e}^{j\angle {{\tilde{S}}}} \approx \textrm{e}^{j\angle {{\tilde{X}}}}$. We define the argument of the  complex number ${\tilde{X}}$ as  $\theta$, then \eqnref{equ:sim2D2} is approximated as
\begin{align}
\label{equ:sim2D3}
{\tilde{Y}}&\approx {{\tilde{S}}}+\textrm{e}^{j\frac{\pi}{2}}{{{Z}}}_p\textrm{e}^{j\theta}= {{\tilde{S}}}+{{{Z}}}_p\textrm{e}^{j(\frac{\pi}{2}+\theta)}\notag\\
&={{\tilde{X}}}+\frac{{{\tilde{W}}}}{\sqrt{P}|\tilde{h}|}+{{{Z}}}_p\textrm{e}^{j(\frac{\pi}{2}+\theta)}.
\end{align}
By convenience, multiplying ${{{\tilde{Y}}}}$ by $\sqrt{P}|\tilde{h}|$ in \eqnref{equ:sim2D3},  the simplified 2D signal model can be rewritten as
\begin{equation}
\label{equ:CDreceiver2D}
{{{\tilde{Y}}}} = \sqrt{P}|\tilde{h}|{{\tilde{X}}}+{{\tilde{W}}}+e^{j(\theta+\frac{\pi}{2})}\frac{{{{Z_r}}}}{\sqrt{\rho}},
\end{equation}
where  ${{{Z_r}}}=\sqrt{\rho P}|\tilde{h}|{{{Z}}}_p$, which denotes the real AWGN with  mean zero and variance $\frac{\sigma_{\textrm{cov}}^2}{2}$. Since
$e^{j(\theta+\frac{\pi}{2})}=-\sin\theta+j\cos\theta$, the 2D signal is further given by
\begin{equation}
\label{equ:CDreceiver2D1}
{{{\tilde{Y}}}} = \sqrt{P}|\tilde{h}|{{\tilde{X}}}+{{{W}}}_r- \sin\theta \frac{{{{Z_r}}}}{\sqrt{\rho}}+j({{{W}}}_i+\cos\theta\frac{{{{Z_r}}}}{\sqrt{\rho}}),
\end{equation}
where $\sin\theta=\frac{{{{X}}}_i}{|\tilde{X}|}$ and $\cos\theta=\frac{{{{X}}}_r}{|\tilde{X}|}$.
Assuming that ${Y}_{r}$ and ${Y}_{i}$ are respectively the real part and imaginary part of the output signal $\tilde{Y}$.  According to the definition of the joint PDF~\cite[(12)]{370165},  the joint PDF of ${Y}_{r}$ and ${Y}_{i}$ for given $\tilde{X}$ is represented as
\begin{align}
\label{equ:2DpdfY1}
f_{{Y}_{r},{Y}_{i}}&({y}_{r},{y}_{i}|\tilde{X})=\frac{1}{2\pi\sigma_{{Y}_{r}}\sigma_{{Y}_{i}}\sqrt{1-\xi^2}}\exp\biggl(-\frac{1}{2({1-\xi^2})}\notag \\
&\biggl(\frac{({y}_{r}-\sqrt{P}|\tilde{h}|{{{X}}}_r)^2}{\sigma_{{Y}_{r}}^2}+\frac{({y}_{i}-\sqrt{P}|\tilde{h}|{{{X}}}_i)^2}{\sigma_{{Y}_{i}}^2}\notag\\
&-\frac{2\xi({y}_{r}-\sqrt{P}|\tilde{h}|{{{X}}}_r)({y}_{i}-\sqrt{P}|\tilde{h}|{{{X}}}_i)}{\sigma_{{Y}_{r}}\sigma_{{Y}_{i}}}\biggr)\biggr),
\end{align}
where $\sigma_{{Y}_{r}}^2=\frac{\sigma_{\textrm{A}}^2}{2}+\frac{\sin^2\theta\sigma_{\textrm{cov}}^2}{2\rho}$ denots the variance of the received signal ${Y}_{r}$, and
 $\sigma_{{Y}_{i}}^2=\frac{\sigma_{\textrm{A}}^2}{2}+\frac{\cos^2\theta\sigma_{\textrm{cov}}^2}{2\rho}$  represents the variance of the received signal ${Y}_{i}$.
 The means of ${Y}_{r}$ and ${Y}_{i}$ are $\sqrt{P}|\tilde{h}|{{{X}}}_r$ and $\sqrt{P}|\tilde{h}|{{\tilde{X}}}_i$, respectively.
 Note that $\xi$ denotes the correlation coefficient, which is calculated as
$\xi=-\frac{\sin\theta\cos\theta\sigma_{\textrm{cov}}^2}{2\rho\sigma_{{Y}_{r}}\sigma_{{Y}_{i}}}$.

\subsection{Low-complexity Signal Detector}
As seen in \eqnref{equ:2DpdfY1}, the variables ${Y}_{r}$ and ${Y}_{i}$ are correlated.  To remove the correlation of the received signal ${Y}_{r}$ and ${Y}_{i}$, we express the joint PDF of the received signal in the new coordinate system.
Since the detection error probability remains invariant for any rotation of the coordinate system, the PDF of the received signal can be represented as another form by rotating the \textit{In-Phase} and \textit{Quadrature} axes~\cite{anton2010elementary}.
In the new coordinate system,  the received signal $({Y}_{r},{Y}_{i})$ is changed as $({Y}_{r}',{Y}_{i}')$, which is given by
\begin{align}
\label{equ:2Dsimnewco}
&{Y}_{r}'= {Y}_{r}\cos\theta+{Y}_{i}\sin\theta-\sqrt{P}|\tilde{h}||\tilde{X}|, \notag\\
&{Y}_{i}'= {Y}_{i}\cos\theta-{Y}_{r}\sin\theta.
\end{align}
Note that ${Y}_{r}'$ and ${Y}_{i}'$ for the \textit{In-Phase} and \textit{Quadrature} branches are independent of each other.  Then the PDF of the received signal given $\tilde{X}$ in \eqnref{equ:2DpdfY1} becomes
\begin{align}
\label{equ:2DpdfY2}
&f_{{Y}_{r}',{Y}_{i}'}({y}_{r}',{y}_{i}'|\tilde{X})=\frac{1}{2\pi\sigma_{{Y}_{r}'}\sigma_{{Y}_{i}'}}\exp\left(-\left(\frac{({y}_{r}')^2}{2\sigma_{{Y}_{r}'}^2}+\frac{({y}_{i}')^2}{2\sigma_{{Y}_{i}'}^2}\right)\right)\notag\\
&\!\!=\!\!\frac{1}{\pi\sqrt{\sigma_{\textrm{A}}^2(\sigma_{\textrm{A}}^2+\frac{\sigma_{\textrm{cov}}^2}{\rho})}}\exp\left(-\left(\frac{({y}_{r}')^2}{\sigma_{\textrm{A}}^2}+\frac{({y}_{i}')^2}{\sigma_{\textrm{A}}^2+\frac{\sigma_{\textrm{cov}}^2}{\rho}}\right)\right),
\end{align}
where $\sigma_{{Y}_{r}'}^2=\frac{\sigma_{\textrm{A}}^2}{2}$ and $\sigma_{{Y}_{i}'}^2=\frac{\sigma_{\textrm{A}}^2}{2}+\frac{\sigma_{\textrm{cov}}^2}{2\rho}$, which means that the noise projections on the  \textit{In-Phase} axis and \textit{Quadrature} axis are $\frac{\sigma_{\textrm{A}}^2}{2}$ and $\frac{\sigma_{\textrm{A}}^2}{2}+\frac{\sigma_{\textrm{cov}}^2}{2\rho}$, respectively.

Based on the received signal PDF $f_{Y'_{r},
Y'_i}\big(y'_r,y'_i|\tilde{X}\big)$ in \eqnref{equ:2DpdfY2}, the low-complexity signal detection rule is expressed as
\begin{align}
\label{equ:MLsub2opt2}
\hat{X}=\mathop{\arg\max} \limits_{\tilde{X}\in \Omega_{\textrm{gen}}} f_{Y'_{r},
Y'_i}\big(y'_r,y'_i|\tilde{X}\big),
\end{align}
where $\Omega_{\textrm{gen}}$ is the set of the transmitted constellation points.
 To detect the transmitted signal $\tilde{X}_{k}$ for the $k$-th symbol, we need to have $f_{Y'_{r},
Y'_i}\big({y'}_r,y'_i|\tilde{X}_{k}\big) > f_{Y'_{r},
Y'_i}\big({y'}_r,y'_i|\tilde{X}_{j}\big)$ for
different $k$ and $j$, i.e.
\begin{align}
\label{equ:MLva}
\exp &\left(-\left(\frac{({y'}_{r}^{(k)})^2}{\sigma_{\textrm{A}}^2}+\frac{({y'}_{i}^{(k)})^2}{\sigma_{\textrm{A}}^2+\frac{\sigma_{\textrm{cov}}^2}{\rho}}\right)\right)>\notag\\
&\exp\left(-\left(\frac{({y'}_{r}^{(j)})^2}{\sigma_{\textrm{A}}^2}+\frac{({y'}_{i}^{(j)})^2}{\sigma_{\textrm{A}}^2+\frac{\sigma_{\textrm{cov}}^2}{\rho}}\right)\right),
\end{align}
where ${y'}_{r}^{(k)}$ and ${y'}_{i}^{(k)}$ denote the received signal when the signal $\tilde{X}_{k}$ transmitted. ${y'}_{r}^{(j)}$ and ${y'}_{i}^{(j)}$ denote the received signal when the signal $\tilde{X}_{j}$ transmitted. Note that $\tilde{X}_j\in \Omega_{\textrm{gen}}$ and $\tilde{X}_j\neq \tilde{X}_k$. Thus, the signal detection is equivalent to the minimum distance detection, which is given by
\begin{align}
\label{equ:MLsub2opt3}
\hat{X}&=\mathop{\arg\min} \limits_{\tilde{X}\in \Omega_{\textrm{gen}}} \left(\frac{({y}_{r}')^2}{\sigma_{\textrm{A}}^2}+\frac{({y}_{i}')^2}{\sigma_{\textrm{A}}^2+\frac{\sigma_{\textrm{cov}}^2}{\rho}} \right)\notag\\
&=\mathop{\arg\min} \limits_{\tilde{X}\in \Omega_{\textrm{gen}}} \Upsilon,
\end{align}
where ${y}_{r}'= {y}_{r}\cos\theta+{y}_{i}\sin\theta-\sqrt{P}|\tilde{h}||\tilde{X}|$,
and ${y}_{i}'= {y}_{i}\cos\theta-{y}_{r}\sin\theta$. Taking ${y}_{r}'$ and ${y}_{i}'$ into \eqnref{equ:MLsub2opt3}, $\Upsilon$ is given by
\begin{align}
\label{equ:MLsub2va2}
\Upsilon=&\frac{\left({y}_{r}\frac{{{{X}}}_r}{|\tilde{X}|}+{y}_{i}\frac{{{{X}}}_i}{|\tilde{X}|}-\sqrt{P}|\tilde{h}||\tilde{X}|\right)^2}{\sigma_{\textrm{A}}^2}\notag\\
&+\frac{\frac{1}{|\tilde{X}|^2}\left({y}_{i}{{{X}}}_r-{y}_{r}{{{X}}}_i\right)^2}{\sigma_{\textrm{A}}^2+\frac{\sigma_{\textrm{cov}}^2}{\rho}}.
\end{align}

According to \eqnref{equ:MLsub2opt3} and \eqnref{equ:MLsub2va2}, for each candidate transmit
symbol $\tilde{X}$, the receiver can determine the detection output that minimizes $\Upsilon$. The detailed steps of signal detector are summarized in Algorithm 1. Notably, this detection scheme is asymptotically optimal as the transmit power $P$ increases. This is due to the fact that the derivation of the low-complexity detector ignores one noise term which has vanishing importance as the SNR increases~\cite{wang2020on}.
\begin{algorithm}[t!]
	\caption{The proposed low-complexity signal detector}
	\begin{algorithmic}[1]
		\State Express the 2D approximated signal $\tilde{Y}$ based on  $({{{\tilde{Y}}}_1}, {{{Y}}_2})$ according to \eqnref{equ:sim2D}.
		\State Simplify the 2D signal $\tilde{Y}$ by the first-order Taylor-series approximation method.
        \State Obtain the 2D simplified signal ${{{\tilde{Y}}}} = \sqrt{P}|\tilde{h}|{{\tilde{X}}}+{{{W}}}_r- \sin\theta \frac{{{{Z_r}}}}{\sqrt{\rho}}+j({{{W}}}_i+\cos\theta\frac{{{{Z_r}}}}{\sqrt{\rho}})$.
		\State Calculate the joint PDF $f_{{Y}_{r},{Y}_{i}}({y}_{r},{y}_{i}|\tilde{X})$ according to \eqnref{equ:2DpdfY1}.
\State  Rotate the coordinates system ${Y}_{r}'= {Y}_{r}\cos\theta+{Y}_{i}\sin\theta-\sqrt{P}|\tilde{h}||\tilde{X}|, {Y}_{i}'= {Y}_{i}\cos\theta-{Y}_{r}\sin\theta$.
\State Express the joint PDF $f_{{Y}_{r}',{Y}_{i}'}({y}_{r}',{y}_{i}'|\tilde{X})$  given in \eqnref{equ:2DpdfY2} in the new coordinate system.
\State Design the low-complexity signal detector based on $f_{{Y}_{r}',{Y}_{i}'}({y}_{r}',{y}_{i}'|\tilde{X})$.
\State Obtain the low-complexity signal detector shown in \eqnref{equ:MLsub2opt3}.
	\end{algorithmic}
\end{algorithm}

\section{Numerical Results}
In this section, we simulate the performance of the proposed signal detection method with perfect channel state information. The antenna noise and post-processing noises  are both considered in the splitting receiver architecture. Since it is difficult to define the SNR for the splitting receiver, the transmit signal power $P$ can also be expressed as the SNR when the noise variances are given, as did in~\cite{Liu2017A}.

\begin{figure}[t]
\centering
\includegraphics[width=0.65 \linewidth]{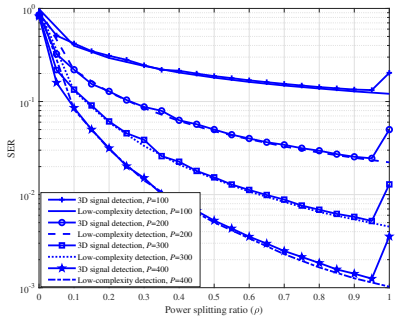}
\caption{SER versus $\rho$ for 64-QAM, $\sigma_{\textrm{{A}}}^2=\sigma_{\textrm{{cov}}}^2=1,\sigma_{\textrm{{rec}}}^2=0.1$.}
\label{fig:64QAMAppandsim}
\end{figure}

\figref{fig:64QAMAppandsim} shows the SER versus the power splitting ratio $\rho$ for 64-QAM with different signal powers. The variances of the antenna noise, conversion noise and  rectifier noise are respectively designed as $\sigma_{\textrm{A}}^2= \sigma_{\textrm{cov}}^2=1$, and $\sigma_{\textrm{rec}}^2=0.1$.
As shown in \figref{fig:64QAMAppandsim}, the approximation SER achieved by the proposed low-complexity signal detection  almost approaches the accurate SER achieved by the 3D signal detection when $\rho\in(0,0.95]$. Hence, the low-complexity signal detection method achieves asymptotically optimal SER performance at a certain power splitting ratio.

\begin{figure}[t]
\centering
\includegraphics[width=0.65 \linewidth]{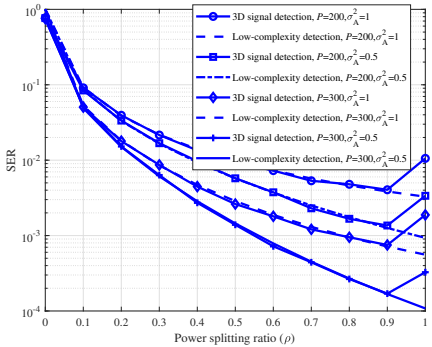}
\caption{SER versus $\rho$ for 32-APSK, $\sigma_{\textrm{{cov}}}^2=\sigma_{\textrm{{rec}}}^2=1$.}
\label{fig:32APSKser}
\end{figure}

Apart from 64-QAM, \figref{fig:32APSKser}  shows the SER versus the power splitting ratio $\rho$  for 32-APSK scheme.  The 32-APSK has four rings with 6, 8, 8, and 10 points on each ring.
In the simulation, the signal powers are respectively designed as 200 and 300. The ring radii are assumed to be ordered and uniformly spaced \cite{Hager2013Design}.  As indicated in \figref{fig:32APSKser}, we see almost no difference between the low-complexity signal detection and  3D signal detection schemes when $\rho\in(0,0.9]$. This
implies that the SER approximation achieved by the proposed low-complexity detection gives nearly optimal performance at a certain power splitting ratio.

\begin{figure}[t]
\centering
\includegraphics[width=0.7 \linewidth]{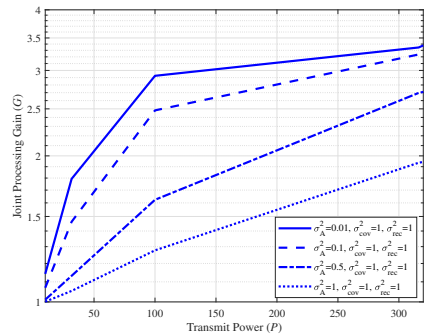}
\caption{Joint processing gain versus $P$ for 64QAM.}
\label{fig:16QAMgains}
\end{figure}

\figref{fig:16QAMgains} demonstrates the joint processing gain in terms of SER versus the signal power $P$. The definition of joint processing gain $G$  is shown in~\cite{Liu2017A}, which is represented as the maximum SER reduction compared to the traditional coherent and non-coherent receivers. As shown in \figref{fig:16QAMgains}, the joint processing gain $G\geq 1$ for different noise variances and signal powers, which means that the splitting receiver scheme obtains an enhanced SER performance compared to the traditional receivers.
%


%
\section{Computational Complexity Analysis}
For the proposed low-complexity signal detection scheme, we transform the 3D signal model into the 2D signal model. As a benefit,  the signal detection is simplified as the minimum distance detection, which calculates the
distance between two signal points. For the traditional CD receiver, the signal detector can also be simplified as the minimum distance detection based on the ML rule \cite{tse2005fundamentals}.
Thus, regarding the computational
complexity,  the proposed signal detector used in the splitting receiver architecture is approximately equal to that of the traditional CD receiver.

Furthermore,  we compare the detection complexity with different signal detectors for the splitting receiver architecture. For the optimal ML detection with 3D received signal in~\cite[(19)]{wang2020on}, it needs to calculate double integrals for achieving the likelihood function. Then the computational complexity of 3D ML detection is extremely high.
The low-complexity detection rule with a closed-form likelihood function was also developed in~\cite{wang2020on}, which can be evaluated quickly compared with the 3D ML detection. However, the obtained closed-form PDF expression in~\cite{wang2020on} is still very complicated, resulting that the detection complexity in~\cite{wang2020on} is much higher than the proposed signal detection method.  Thus, the proposed signal detector greatly reduces the
computational complexity of signal detection at the receiver.

To further illustrate the advantage of our proposed signal detection scheme, we quantitatively analyze the computational complexity of the proposed signal detection and the method
in~\cite{wang2020on}. Actually, the calculation of the PDF value in~\cite[(26)]{wang2020on} requires an exponential operation and about $300$ multiplication operations. It is shown in~\cite{600828} that the computational complexity of the exponentiation $\exp(\cdot)$ is equivalent to $2\log_2(\cdot)$ numbers of multiplication operations. Therefore, we need approximately $2 \log_2 (\cdot)+300$ multiplications to compute one PDF in~\cite[(26)]{wang2020on}.
Whereas, the computation of \eqnref{equ:MLsub2va2} only requires about $27$ multiplications. Thus, the computational complexity of the proposed scheme is at least reduced by about  $11.1$  times.

As for the hardware complexity, the splitting receiver
is also easily implemented by adding a PD receiver to the CD receiver as depicted in \figref{fig:Fig1sysmod}, where the most commonly used PD circuits are passive~\cite{Liu2017A, WangGCJ21}. Although the splitting
receiver architecture is slightly more complex than that of the traditional CD
receiver,  the splitting receiver ($0<\rho<1$) achieves lower
SER than the CD receiver ($\rho=1$), which has been shown in Section IV.
\section{Conclusions}
This paper has investigated the signal detection scheme for the splitting receiver architecture with consideration of practical noise model. Based on the derived two-dimensional signal model, a low-complexity signal detection method has been proposed. We have shown
that the computational complexity of our proposed signal detection scheme can be  reduced by about $11.1$ times compared to the method in~\cite{wang2020on}.
Simulation results have confirmed that the approximation SER achieved by the proposed low-complexity signal detection method  approaches the accurate SER at a certain splitting ratio.

\bibliographystyle{IEEEtran}
\bibliography{ref}

\end{document}